\documentclass{article}
\usepackage[utf8]{inputenc}
\usepackage[T1]{fontenc}

\usepackage[affil-it]{authblk}
\usepackage{graphicx}
\usepackage{hyperref}

\usepackage{tikz}
\usetikzlibrary{patterns}
\usepackage{subfig}
\usepackage{amsmath}
\usepackage{mathtools}
\usepackage{tabularx}

\usepackage{xspace}
\usepackage{booktabs}

\newcommand{\N}{{\cal N}}
\newcommand{\M}{{\cal M}}
\newcommand{\T}{{\cal T}}
\newcommand{\F}{{\cal F}}
\newcommand{\nl}{\tabularnewline}
\newcolumntype{C}{>{\centering}X}
\newcolumntype{P}[1]{>{\centering\arraybackslash}p{#1}}
 \newtheorem{example}{Example}
 \newtheorem{remark}{Remark}

\newcommand{\IP}{ILP model\xspace}
\newcommand{\CP}{CP$_{O}$ model\xspace}
\newcommand{\CPN}{CP$_{N}$ model\xspace}
\newcommand{\FQ}{$(flow-qual)$\xspace}
\newcommand{\QF}{$lex(-qual,flow)$\xspace}
\newcommand{\ttit}[1]{\multicolumn{1}{c}{#1}}

\allowdisplaybreaks[4]

\begin{document}

\title{A new CP-approach for a parallel
  machine scheduling problem with time constraints on machine qualifications.}

\author{Arnaud Malapert}
\affil{ Université Côte d'Azur, CNRS, I3S, France \\
 \texttt{arnaud.malapert@unice.fr}} 
\author{Margaux Nattaf}
\affil{Univ. Grenoble Alpes, CNRS, Grenoble INP, G-SCOP, 38000 Grenoble, France\\ 
\texttt{margaux.nattaf@grenoble-inp.fr}}


  % 
\maketitle              % typeset the header of the contribution

\begin{abstract} 
  This paper considers the scheduling of job families on parallel
  machines with time constraints on machine qualifications. In this
  problem, each job belongs to a family and a family can only be
  executed on a subset of qualified machines. In addition, machines can
  lose their qualifications during the schedule. Indeed, if no job of a
  family is scheduled on a machine during a given amount of time, the
  machine lose its qualification for this family. The goal is to
  minimize  the sum  of job completion times,  i.e. the  flow time,
  while maximizing the number of qualifications 
  at the end of the schedule. The paper presents a new Constraint
  Programming model taking more advantages of the CP feature to model
  machine disqualifications. This model is compared with two existing
  models: an Integer Linear Programming (ILP) model and a Constraint
  Programming (CP) model. The experiments show that the new CP model
  outperforms the other model when the priority is given to the number
  of disqualifications objective. Furthermore, it is competitive with the
  other model when the flow time objective is prioritized.

  \bigskip
  
 \textbf{keywords : }Parallel Machine Scheduling, Time Constraint, Machine
 Qualifications, Integer Linear Programming, Constraint Programming.
\end{abstract}

\section{Introduction}
\vspace{-0.1cm}

Process industries, and specially semiconductor industries, need to be
more and more competitive and they are looking for strategies to
improve their productivity, decrease their costs and enhance
quality. In this context, companies must pay constant attention to
manufacturing processes, establish better and more intelligent
controls at various steps of the fabrication process and develop new
scheduling techniques. One way of doing it is to integrate scheduling
and process control~\cite{Yugma2015}. This paper considers such a
problem: the integration of constraints coming from process control
into a scheduling problem.

Semiconductor fabrication plants (or fabs) have characteristics that
make scheduling a very complex issue~\cite{Moench2011}. Typical ones
includes a very large number of jobs/machines, multiple job/machine
types, hundreds of processing steps, re-entrant flows, frequent
breakdowns... Scheduling all jobs in a fab is so complex that jobs are
scheduled in each workshop separately. In this paper, the focus is on
the photolithography workshop, which is generally a bottleneck
area. In this area, scheduling can be seen as a scheduling problem on
non-identical parallel machines with job family setups (also called
s-batching in~\cite{Moench2011}).

Fabrication processes of semiconductors are very precise and require a
high level of accuracy. Reliable equipment are required and valid 
recipe parameters should be provided. Advanced Process Control (APC)
systems ensure that each process is done following predefined
specifications and that each equipment is reliable to process
different product types. APC is usually associated with the
combination of Statistical Process Control, Fault Detection and
Classification, Run to Run (R2R) control, and more recently
Virtual Metrology~\cite{Moyne2000}. The main interest of this
paper is to consider, in scheduling decisions, constraints induced by
R2R controllers. As shown in the survey paper of~\cite{Tan2015}, R2R
control is becoming critical in high-mix semiconductor manufacturing
processes.

R2R controller uses data from past process runs to adjust settings for
the  next   run  as  presented  for   example  in~\cite{Musacchio1997}
and~\cite{Jedidi2011}. Note that a R2R controller is associated to one
machine  and one  job  family. In  order to  keep  the R2R  parameters
updated  and   valid,  a  R2R   control  loop  should   regularly  get
data.   Hence,   as  presented   in~\cite{Nattaf2018R,Obeid2014},   an
additional constraint is  defined on the scheduling  problem to impose
that the execution of two jobs of  the same family lies within a given
time interval on  the same (qualified) machine. The value  of the time
threshold  depends  on  several  criteria such  as  the  process  type
(critical or  not), the equipment  type, the stability of  the control
loop, etc. If  this time constraint is not  satisfied, a qualification
run is required  for the machine to   be able to process again the job
family on the  machine. This procedure ensures that  the machine works
within a specified  tolerance and is usually  time-consuming.  In this
paper,  we  assume that  qualification  procedures  are not  scheduled
either  because the  scheduling  horizon is  not  suficiently long  or
because qualification procedures have  to be manually performed and/or
validated  by  process   engineers.   Therefore,  maintaining  machine
qualifications as long as possible  is crucial.  More precisely, it is
important to have as many remaining machine qualifications as possible
at the end of the schedule, so that future jobs can also be scheduled.

To our knowledge, there are few articles dealing with scheduling
decisions while integrating R2R constraints. \cite{Cai2011}
and~\cite{Li2008} study related problems, except that they allow
qualification procedures to be performed, the number or the type of
machines is different and the threshold is expressed in number of jobs
instead of in time. The scheduling problem addressed in this paper has
been introduced in~\cite{Obeid2014}, where two Integer Linear Programs
(ILP) and two constructive heuristics are proposed. More
recently,~\cite{Nattaf2018C,Nattaf2018R} develop a new ILP, modelling
problem constraints in a better way. The paper also present one
Constraint Programming (CP) model, as well as two improvement
procedures of existing heuristics.

In this paper, a new CP model, which take more advantages of the CP
features, is presented. The main idea of this model is to exploit the
fact that once a machine is disqualified, it is until the end of the
schedule. The consequence of this is that it is possible to model
machine disqualifications more accurately.  Then, the performance of
this model is compared with the two exact solution methods described
in~\cite{Nattaf2018C}. The paper is organized as
follows. Section~\ref{sec:pb} gives a formal description of the
problem. Section~\ref{sec:ILPCP} presents the two models
of~\cite{Nattaf2018C}. Section~\ref{sec:newCP} describes the new CP
model and finally, Section~\ref{sec:expe} provides a detailed
comparison of the performance of each model.

\vspace{-0.1cm}

\section{Problem description}
\label{sec:pb}

\vspace{-0.1cm}

Formally, the problem takes as input a set of jobs,
$\N=\{1,\dots,N\}$, a set of families $\F =\{1,\dots, F\}$ and a set
of machines, $\M=\{1,\dots,M\}$. Each job $j$ belongs to a family and the
family associated with $j$ is denoted by $f(j)$. For each family $f$,
only a subset of machine, $\M_f \subseteq \M$, is able to process
a job of $f$. A machine $m$ is said to be qualified to process a family
$f$ if $m \in \M_f$.

\noindent
Each  family $f$ is associated with the following
parameters: 
\vspace{-0.3cm}
\begin{itemize}
\item $n_f$ denotes the number of jobs in the family. Note that
  $\sum_{f \in\F} n_f = N$.
\item $p_f$ corresponds to the processing time of jobs in $f$.
\item $s_f$ is the setup time required to switch the production from
  a job belonging to a family $f' \neq f$ to the execution of a job
  of $f$. Note that this setup time is independent of $f'$. In
  addition, no setup time is required between the execution of two
  jobs of the same family. 
\item $\gamma_f$ is the threshold value for the time interval between
the execution of two jobs of $f$ on the same machine. Note that this
time interval is computed on a start-to-start basis, i.e. the
threshold is counted from the start of a job of family $f$ to the
start of the next job of $f$ on machine $m$. Then, if there is a time
interval $]t,t +\gamma_f]$ without any job of $f$ on a machine, the
machine lose its qualification for $f$. 
\end{itemize}

The objective  is to minimize  both the  sum of job  completion times,
i.e.   the  flow time,  and  the  number  of qualification  looses  or
disqualifications.  Note that the interest of minimizing the number of
disqualifications comes from  the fact that, even if  the time horizon
considered is  relatively small,  the problem is  solved in  a rolling
horizon.  Hence, it is  interesting to preserve machine qualifications
for  future jobs.   In addition,  it is  relevant to  consider that  a
machine cannot  lose its qualification for  a family after the  end of
the schedule.  Thus,  this assumption is made in the  remaining of the
paper.  This problem,  introduced in~\cite{Obeid2014},  is called  the
scheduling Problem  with Time  Constraints (PTC).   An example  of PTC
together with two feasible solutions is now presented.

\begin{example}
  \label{ex:PTC}
  ~\\
  \vspace{-0.7cm}
  
  \begin{minipage}{0.45\linewidth}
  Consider the following instance with $N=10,\ M=2$ and $F=3$:
\end{minipage}
\hfill
\begin{minipage}{0.45\linewidth}
  \vspace{-0.1cm}
  \begin{center}
    \begin{tabularx}{\linewidth}{|c|CCCCC|}
      \hline
      $f$ & $n_f$ & $p_f$ & $s_f$ & $\gamma_f$ & $ \M_f$\nl
      \hline
      1&3&9&1&25&$\{2\}$\nl
      2&3&6&1&26&$\{1,2\}$\nl
      3&4&1&1&21&$\{1,2\}$\nl
      \hline
    \end{tabularx}
  \end{center}
\end{minipage}

\vspace{0.1cm}
Figure~\ref{fig:exPTC} shows two feasible solutions. The first
solution, described by Figure~\ref{subfig:exOptCf}, is optimal in
terms of flow time. For this solution, the flow time is equal to
$1+2+9+15+21+1+2+12+21+30 = 114$ and the number of qualification
losses is $3$. Indeed, machine $1$ ($m_1$) loses its qualification for
$f_3$ at time $22$ since there is no job of $f_3$ starting in interval
$]1,22]$ which is of size $\gamma_3 = 21$. The same goes for $m_2$ and
$f_3$ at time $22$ and for $m_2$ and $f_2$ at time $26$.

The second solution, described by Figure~\ref{subfig:exOptDisq}, is
optimal in terms of number of disqualifications. Indeed, in this
solution, none of the machines loses their qualifications. However,
the flow time is equal to $1+2+9+17+19+9+18+20+27+37=159$. This shows
that the flow time and the number of qualification losses are two
conflicting criteria. Indeed, to maintain machine qualifications, one
need to regularly change the job family executed on machines. This
results in many setup time and then to a large flow time
value.

\vspace{-0.6cm}
  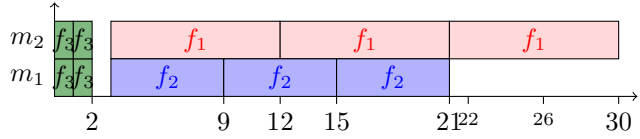
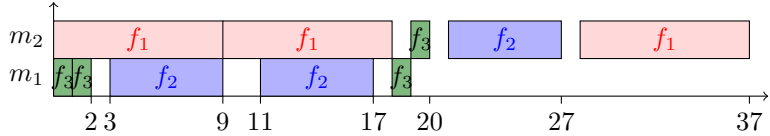
\begin{figure}
    \centering
    \subfloat[An optimal solution for the flow time objective \label{subfig:exOptCf}]{
      \begin{tikzpicture}[xscale=0.25]
        \node (O) at (0,0) {};
        \draw[->] (O.center) -- ( 31,0); 
        \draw[->] (O.center) -- (0,1.25); 
        \draw (0, 0.25) node[left] {$m_1$} ;
        \draw (0, 0.75) node[left] {$m_2$} ;
        \draw (2,0) -- (2,-0.1) node[below] {$2$} ;
        \draw (9,0) -- (9,-0.1) node[below] {$9$} ;
        \draw (12,0) -- (12,-0.1) node[below] {$12$} ;
        \draw (15,0) -- (15,-0.1) node[below] {$15$} ;
        \draw (21,0) -- (21,-0.1) node[below] {$21$} ;
        \draw (22,0) -- (22,-0.1) node[below] {\scriptsize $22$} ;
        \draw (26,0) -- (26,-0.1) node[below] {\scriptsize $26$} ;
        \draw (30,0) -- (30,-0.1) node[below] {$30$} ;
        \draw[fill=red!15!]  (3,0.5) rectangle (12,1)
        node[midway,red] {$f_1$}; 
        \draw[fill=red!15!]  (12,0.5) rectangle (21,1)
        node[midway,red] {$f_1$}; 
       \draw[fill=red!15!]    (21,0.5) rectangle (30,1)
        node[midway,red] {$f_1$}; 
        \draw[fill=blue!30!]  (3,0) rectangle (9,0.5)
        node[midway,blue] {$f_2$}; 
        \draw[fill=blue!30!]  (9,0) rectangle (15,0.5)
        node[midway,blue] {$f_2$}; 
        \draw[fill=blue!30!]  (15,0) rectangle (21,0.5)
        node[midway,blue] {$f_2$}; 
        \draw[fill=green!45!black!50!] (0,0) rectangle (1,0.5) node[midway,black] {$f_3$};
        \draw[fill=green!45!black!50!]  (1,0) rectangle (2,0.5) node[midway] {$f_3$};
        \draw[fill=green!45!black!50!]  (0,0.5) rectangle (1,1) node[midway] {$f_3$};
        \draw[fill=green!45!black!50!]  (1,0.5) rectangle (2,1) node[midway] {$f_3$};
      \end{tikzpicture}
    }
    
    \subfloat[An optimal solution for qualification losses \label{subfig:exOptDisq}]{
      \begin{tikzpicture}[xscale=0.25]
        \node (O) at (0,0) {};
        \draw[->] (O.center) -- ( 38,0);
        \draw[->] (O.center) -- (0, 1.25);
        \draw (0, 0.25) node[left] {$m_1$} ;
        \draw (0, 0.75) node[left] {$m_2$} ;
        
        \draw (2,0) -- (2,-0.1) node[below] {$2$} ;
        \draw (3,0) -- (3,-0.1) node[below] {$3$} ;
        \draw (9,0) -- (9,-0.1) node[below] {$9$} ;
        \draw (11,0) -- (11,-0.1) node[below] {$11$} ;
        \draw (17,0) -- (17,-0.1) node[below] {$17$} ;
        \draw (20,0) -- (20,-0.1) node[below] {$20$} ;
        \draw (27,0) -- (27,-0.1) node[below] {$27$} ;
        \draw (37,0) -- (37,-0.1) node[below] {$37$} ;
        \draw[fill=red!15!]  (0,0.5) rectangle (9,1)
        node[midway,red] {$f_1$}; 
        \draw[fill=red!15!]  (9,0.5) rectangle (18,1)
        node[midway,red] {$f_1$}; 
        \draw[fill=red!15!]  (28,0.5) rectangle (37,1)
        node[midway,red] {$f_1$}; 
        \draw[fill=blue!30!]  (3,0) rectangle (9,0.5)
        node[midway,blue] {$f_2$}; 
        \draw[fill=blue!30!]  (11,0) rectangle (17,0.5)
        node[midway,blue] {$f_2$}; 
        \draw[fill=blue!30!] (21,0.5) rectangle (27,1)
        node[midway,blue] {$f_2$}; 
        \draw[fill=green!45!black!50!]  (0,0) rectangle (1,0.5) node[midway] {$f_3$};
        \draw[fill=green!45!black!50!]  (1,0) rectangle (2,0.5) node[midway] {$f_3$};
        \draw[fill=green!45!black!50!]  (18,0) rectangle (19,0.5) node[midway] {$f_3$};
        \draw[fill=green!45!black!50!]  (19,0.5) rectangle (20,1) node[midway] {$f_3$};
      \end{tikzpicture}
    }
    \caption{Two solution examples for PTC.}
    \label{fig:exPTC}
  
    \vspace{-0.6cm}
  \end{figure}

  Note also that disqualifications may occur after the last job on the
machine.  For example,  in  Figure~\ref{subfig:exOptCf}, $m_1$  become
disqualified for $f_3$ at time $22$  whereas the last job scheduled on
$m_1$ finishes at  time $21$. However, no  disqualifications can occur
after  the makespan $C_{max}$.
\end{example}

\begin{remark}[Bound on the makespan]
  \label{rem:boundCmax}
  In the remaining of the paper, the following upper and lower bound
  on the makespan are defined.  The upper bound used is the same as
  in~\cite{Obeid2014}, i.e. $T=\overline{C_{max}}=\sum_{f \in \F} n_f
  \cdot ( p_f + s_f)$.  A trivial lower bound is $\underline{C_{max}}=
  \lceil (\sum_{f \in \F} n_f * p_f) /M \rceil$. 
\end{remark}

\begin{remark}[Bi-objective optimization]  
 In~\cite{Obeid2014}, PTC is studied using  a weighted sum of the flow
time and  number of disqualifications.  The weight associated  to the
flow  time  is $\alpha$  and  is  always  equal  to $1$.   The  weight
associated with the number of  disqualifications is $\beta$ and is set
to $1$ when the priority is given to  the flow time and to $N \cdot T$
when the priority is given to the number of disqualifications.
In this  paper, two objective  modelling are considered:  the weighted
sum and the lexicographical order. The weighted sum is used in the ILP
model in all cases  and in the CP model only  when the minimization of
the flow time is prioritized. The lexicographical order is used in the
CP   model   when  the   priority   is   given   to  the   number   of
disqualifications.
\end{remark}

\vspace{-0.1cm}

\section{Existing Models} 
\label{sec:ILPCP}

\vspace{-0.1cm}

This   section   describes   the   two   exact   methods   developped
in~\cite{Nattaf2018C}. First,  the ILP is  described and then,  the CP
model. Note  that both  CP model  described in this  paper use  the CP
Optimizer (CPO) framework.  Indeed, CPO allow us to handle in an
efficient way precedence constraints  and optional jobs.  Furthermore,
constraint propagation of these type  of constraints is very efficient
with CPO.

\subsection{ILP model}
\label{sec:ILP}

The ILP model in~\cite{Nattaf2018C} is an improvement of two
existing models  introduced in~\cite{Obeid2014}.  The first  ILP model
of~\cite{Obeid2014} relies on a job-based formulation. Indeed, in this
model,  a variable  $x_{j,t}^m$  is  defined for  each  job $j$,  each
machine $m$ and each  time $t$. This variable is then  equal to $1$ if
and only if job  $j$ starts at time $t$ on machine  $m$. However, in a
solution, there  is no need to  know which job  start at which
time  on  which  machine.  Indeed,  only the  family  of  the  job  is
important.    Hence,    a     family-based    model    is    developed
in~\cite{Obeid2014}(IP2) and improved in~\cite{Nattaf2018C}(IP3).

In the family-based model, a variable $x_{f,t}^m$ is introduced for
each family $f \in \F$, each machine $m \in \M_f$ and each time $t \in
\T=\{0,\dots,T-1\}$, with  $T$ the  upper bound  on the  makespan (see
Remark~\ref{rem:boundCmax}). This variable  is set to one  if and only
if one job  of $f$ starts at  time $t$ on machine  $m$. Therefore, the
number of binary variables is reduced compared to the job-based model.

Similarly,  a   set  of   variable  $y_{f,t}^m$   is  used   to  model
disqualifications. This variable is set to $1$ only if family $f$ lose 
its qualification on machine $m$ at time $t$. However, in (IP2), it
may occur that a machine become disqualified after $C_{max}$.  Thus,
in (IP3), another variable set $Y_f^m$ is defined to model the fact
that a machine becomes disqualified for a family before $C_{max}$.
\footnotesize
  \begin{align}
    & \text{min. } \alpha\cdot \sum_{f \in {\F}} C_f + \beta\cdot\sum_{f
    \in {\F}} \sum_{m \in {\M}} Y_f^m
    \label{eq:objIP3}\\[-4pt]
    & \sum_{m \in \M_f} \sum_{t=0}^{T-p_f}x_{f,t}^m = n_f
    &\forall f \in {\F} \label{eq:nfIP3}\\[-6pt]
    & \sum_{m \in \M_f} \sum_{t=0}^{T-p_f}(t+p_f)\cdot x_{f,t}^m \le
      C_f & \forall f \in {\F}\label{eq:CfIP3}\\[-6pt]
      &  n_f \cdot x_{f',t}^m + \sum_{\tau = t - p_f -
    s_{f'}+1}^t x_{f,\tau}^m \le n_f &  \forall f \neq f' \in {\cal
                                       F}^2,\nonumber \\[-12pt]
    \vspace{-0.2cm}
  & &\mathllap{ \ \forall m \in \M_f
      \cap \M_{f'}, \ \forall t \in \T}
      \label{eq:noOverlapIP3}\\[-5pt]
    & y_{f,t}^m + \sum_{\tau = t - p_f + 1}^{t} x_{f,\tau}^m \le 1 &
\forall f \in {\cal F},\ \forall t \in \T,\ \forall m \in \M_f
                         \label{eq:noOverlapQualifIP3} \\[-6pt]
  &  y_{f,t}^m + \sum_{\tau = t - \gamma_f + 1}^t
    x_{f,\tau}^m \ge 1 & \forall f \in {\cal F},\ \forall t \ge
                         \gamma_f \in \T,\ \forall m \in \M_f
                         \label{eq:disqualifIP3} \\[-2pt]
  &  y_{f,t-1}^m \le y_{f,t}^m & \forall f \in
                                 {\cal F},\ \forall t \in \T,\ \forall m \in \M_f
                                 \label{eq:maintainDisqualifIP3} \\[-2pt]
   \mathrlap{ \frac{1}{M\cdot (T-t)} \sum_{f' \in
    {\cal F}}\sum_{\tau = t - p_{f^{'}}}^ {T-1} \sum_{m' \in \M_{f'}}
    x_{f',\tau}^{m'} + y_{f,t-1}^m - 1 \le Y_f^m } \nonumber \\[-10pt]
    & & \forall t \in \T,\
                                 \forall f \in {\cal F},\ \forall m \in \M_f
                                 \label{eq:aggregatedDisqualifIP3} \\[-2pt] 
  & x_{f,t}^m \in \{0,1\} &\forall t \in \T, \forall f \in
                            {\cal F},\ \forall m \in \M_f \label{eq:boundxIP3}\\[-2pt]
  & y_{f,t}^m \in \{0,1\} & \forall t \in \T, \forall f
                            \in {\cal F},\ \forall m \in \M_f\label{eq:boundyIP3}\\[-2pt]
  & Y_{f}^m \in \{0,1\} & \forall f \in {\cal F},\ \forall m \in
                          \M_f \label{eq:boundYIP3}
\end{align}
\normalsize

The objective of  the model is described  by~\eqref{eq:objIP3}.  It is
expressed  as the  weighted sum  of the  flow time  and the  number of
disqualifications. Constraints~\eqref{eq:nfIP3}  ensure that  all jobs
are  executed. Constraints~\eqref{eq:CfIP3}  is  used  to compute  the
completion time of family $f$, i.e.  the sum of completion time of all
jobs of $f$.  Constraints~\eqref{eq:noOverlapIP3}  ensure that jobs of
$f$ and  jobs of $f'$  does not overlap and  that the setup  times are
satisfied. Constraints~\eqref{eq:noOverlapQualifIP3} are used to model
both the fact that the execution of two jobs of the same family cannot
occur simultaneously and  the fact that a machine has  to be qualified
to process  a job. Constraints~\eqref{eq:disqualifIP3} make  sure that
if  no jobs  of family  $f$  start on  $m$  during an  interval $]t  -
\gamma_f ,  t]$, then $m$  becomes disqualified  for $f$ at  time $t$.
Constraints~\eqref{eq:maintainDisqualifIP3}        maintain        the
disqualification  of   the  machine  once  it   becomes  disqualified.
Finally, Constraints~\eqref{eq:aggregatedDisqualifIP3}  ensure that it
is no  longer necessary to  maintain a  qualification on a  machine if
there is no  job which starts on  any machine in the  remainder of the
horizon, i.e. $\frac{1}{M\cdot (T-t)} \sum_{f' \in {\cal F}}\sum_{\tau
= t - p_{f^{'}}}^ {T-1} \sum_{m' \in \M_{f'}} x_{f',\tau}^{m'} = 0 $.

The number  of variables of the  model is $F\cdot M  \cdot (2T+1)$ and
the number of constraints is at most $2F + T \cdot M \cdot (4F+F^2)$.

\subsection{CP model}
\label{sec:CP}

In  this   part,  the   CP  model  defined   in~\cite{Nattaf2018C}  is
described. The  first part of  the model  concerns the modelling  of a
classical parallel  machine scheduling problem (PMSP)  with setup time
and the second part deals with the modelling of the disqualifications.
%More attention will be given to this part of the model.

\vspace{-0.3cm}

\subsubsection{The  parallel  machine  scheduling problem  with  setup
  time} 
The  PMSP with  setup  time can  be modeled  using  optional (or  not)
interval   variables    introduced  by~\cite{Laborie08,Laborie09}.  An
(optional) interval variable $J$ is  associated with four variables: a
start time,  $st(J)$; a duration, $d(J)$;  an end time, $et(J)$  and a
binary  execution status  $x(J)$,  equal to  $1$ if  and  only if  the
interval variable is present in the  final solution. If the job $J$ is
executed, it behaves  as a classical job that is  executed on its time
interval, otherwise it is not considered by any constraint.

In the considered  scheduling problem, a job $j$ of  family $f$ can be
scheduled  on any  machine belonging  to $\M_f$.  Therefore, a  set of
optional interval  variables $altJ_{j,m}$ is associated  with each job
$j$  and each  machine belonging  to $\M_{f(j)}$.  The domain  of such
variables  is $dom(altJ_{j,m})=  \{  [st,et)\,  |\, [st,et)  \subseteq
[0,T),\ st  + p_{f(j)} =  et\}$. Furthermore, a  non-optional interval
variable,  $jobs_j$ is  associated with  each job  $j$. Its  domain is
$dom(jobs_{j})=  \{  [st,et)\,  |\,  [st,et) \subseteq  [0,T),\  st  +
p_{f(j)} = et\}$.

To model  the PMSP with setup  time, the following two  sets of global
constraints is used~\cite{Wolf2009}.

\paragraph{Alternative constraints}

Introduced  in~\cite{Laborie08}, this  constraint models  an exclusive
alternative between a bunch of jobs.
  \footnotesize
\begin{align}
  alternative \left(jobs_j, \left\{ altJ_{j,m} | m \in \M_{f(j)} \right\}
  \right) & &  \forall j \in \N
\end{align}
\normalsize
It  means that  when $jobs_j$  is executed,  then exactly  one of  the
$altJ_{j,m}$ jobs must be executed,  i.e. the one corresponding to the
machine $m$ on which the job is scheduled. Furthermore, the start date
and the end  date of $jobs_j$ must be synchronised  with the start and
end  date  of the  $altJ_{j,m}$  jobs.  However,  if $jobs_j$  is  not
executed,  none of  the  other jobs  can be  executed.  In our  model,
$jobs_j$ is a mandatory job. This constraint models the fact that each
job must be executed on one and only one machine.

\paragraph{No-Overlap constraints}

An important constraint is that jobs cannot be executed simultaneously
on the same  machine. It is a unary resource  constraint. Each machine
can then be used  by only on job at a time. To  model this feature, we
use noOverlap constraints. This  constraint ensures that the execution
of several interval  variables do not overlap. It can  also handle the
setup time. Let $S$ be the matrix of setup times of the problem, i.e.
$\left(S_{f',f}\right) = \left\{
  \begin{array}{lcl}
    0 & & \text{if } f = f' \\
    s_f & & \text{oherwise}
  \end{array}
\right.$.
Then,  the following  noOverlap constraint  makes sure  that, for  all
pairs of jobs $(i,  j)$ s.t. $m \in \M_i \cap  \M_j$, either the start
of $altJ_{j,m}$ occurs  after the end of  $altJ_{i,m}$ plus $s_{f(j)}$
or the opposite:
  \footnotesize
\begin{align}
  noOverlap  \left(  \left\{  altJ_{j,m}   |\forall  j\  s.t.\  m  \in
  \M_{f(j)} \right\}, S \right) 
  & &  \forall m \in \M
\end{align}
\normalsize
The exact semantic of this constraint is presented
in~\cite{Laborie09}.

\paragraph{Additional ordering constraints}
The authors of~\cite{Nattaf2018C} add a non-mandatory set of
constraints to the model. Indeed, the model is correct without these
constraints but adding them remove many symmetry in the model. The
constraint order the start of jobs belonging to the same family.
\footnotesize
\begin{align}
  startBeforeStart(jobs_j,jobs_j') & & \forall j,j' \in \N,\ j> j', f(j')=f(j)
\end{align}
\normalsize

\subsubsection{Modelling of the number of disqualifications}

In the model of~\cite{Nattaf2018C}, disqualifications are modelled as
optional interval variables. The variable will be present in the final
solution if and only if, the machine became disqualified for the
family. The start time of the variable corresponds to the time at
which the machine becomes disqualified. Therefore, a set of optional
interval variable, $disq_{f,m}$, of length $0$ is defined for each family
$f$ and each machine $m$ such that  $m \in \M_f$.  The domain of these
variables  is  $dom(disq_{f,m})  =  \{ [st,et)\,  |\,  [st,et)  \subseteq
[\gamma_f  , T),\  st  = et\}$.  In  addition, the  model  will use  a
$C_{max}$ interval  variable of length  $0$ modelling the end  time of
the  last  job   executed  on  all  machine,  i.e.  the   end  of  the
schedule.  Its  domain  is  $dom(C_{max})= \{  [st,et)\,  |\,  [st,et)
\subseteq [0 , T),\ st = et\}$. 

Then, the constraints used to model machine disqualifications are
stated below. The first set of constraints model the fact that each
job has to be executed before $C_{max}$.
\footnotesize
\begin{align}
  endBeforeStart(jobs_j, C_{max}) & &\forall j \in \N 
\end{align}
\normalsize
Another set of constraints ensures that no job of a family $f$ is
scheduled on $m$ if $m$ is disqualified for $f$, i.e. after the
$disq_{f,m}$ job.
\footnotesize
\begin{align}
  startBeforeStart(altJ_{j,m},disq_{f(j),m},\gamma_{f(j)}) & & \forall
  j \in \N, \ \forall m \in \M_{f(j)} 
\end{align}
\normalsize
Finally, the following constraints sets enforce a machine to become
disqualified if no job of family $f$ is scheduled on $m$ during an
interval of size $\gamma_f$. Indeed, the first set state that if a job
of $f$ is scheduled on $m$, either there is another job of $f$
scheduled on $m$ less than $\gamma_f$ units of time later, or the
machine become disqualified, or the end of the scheduled ($C_{max}$)
is reached.
\footnotesize
\begin{align} 
  & x(altJ_{j,m}) \Rightarrow
    \hspace{-0.4cm} \bigvee_{\substack{j' \neq j \\f(j) = f(j')}}
  \hspace{-0.4cm}
  \left( st(altJ_{j',m}) \le t_{j,m} \right) \vee
  \left( st(disq_{f(j),m}) = t_{j,m} \right) \vee
  \left(C_{max} \le t_{j,m} \right) & \nonumber \\[-12pt]
  & &\mathllap{ \forall j \in \N,\  \forall m \in  \M_{f(j)}~~~}
\end{align}
\normalsize
 \noindent  with  $t_{j,m} =  st(altJ_{j,m})  +  \gamma_{f(j)} $.  The
 second set of constraints ensures that  if no job of $f$ is scheduled
 on $m$, then $m$ becomes disqualified for $f$. 
\footnotesize
\begin{align}
& \hspace{-1cm} \bigvee_{\substack{j \in \N\\f(j) =
  f(j')}} \hspace{-0.4cm} \left( st(altJ_{j,m})  \le  \gamma_f
  \right) \vee 
  \left( st(disq_{f(j),m}) = \gamma_f \right) \vee   \left(
  st(C_{max})  \le  \gamma_f \right) 
                                          &   \nonumber \\[-12pt]
  & & \mathllap{ \forall f \in  \F,
                                         \forall  m    \in    \M_f}\hspace{-0.5cm} 
\end{align}
\normalsize

\vspace{-0.2cm}
\subsubsection{Objective function}

In the considered problem, the objective is to minimize both the
flow-time and the number of disqualifications. In this CP model, the
flow-time can be expressed as $flowTime = \sum_{j \in \N} et(jobs_j)$
and the number of disqualifications as $nbDisq= \sum_{f \in \F}
\sum_{m \in \M} x(disq_{f,m})$.

\vspace{-0.2cm}
\subsubsection{Model size}

The number of variables of the model is at most $N\cdot (M + 1) + M\cdot F + 1$ and
the number of constraints is at most $N^2 + 2N + M \cdot (1 + 2N + F)$.

\section{New CP Model}
\label{sec:newCP}

This section presents a new CP model that can be used to solve
PTC. As said earlier, PTC can be decomposed into two sub-problems: a
PMSP with setup time and a {\it machine qualifications problem}. The
model described in this section uses the same idea as 
in~\cite{Nattaf2018C} to formulate the first sub-problem of
PTC. However, to model the machine qualification sub-problem a novel
approach is developed modelling qualifications as resource.

The first part of this section described the difference of modelling
of the PMSP between the model of Section~\ref{sec:CP} and the model of
this section. The second part is dedicated to the machine
qualification sub-problem.

In the model, the two following assumptions are made. First, it is
assumed that jobs of the same family have consecutive index in
$\N$. More precisely, with $n_f$ the number of jobs in family $f$ then
jobs with index in $\{1,\dots,n_1\}$ belong to family $1$, jobs with
index in $\{n_1+1,\dots,n_1+n_2\}$ are jobs of family $2$, etc.  The
second assumption made in the model is that it is equivalent to
consider the threshold either on an end-to-end basis or on a
start-to-start basis. Indeed, if a job of family $f$ starts at time
$t$ on $m$, another job of $f$ has to start before $t+\gamma_f$. This
is equivalent to: if a job of family $f$ ends at time $t+p_f$ on $m$,
another job of $f$ has to end before $t+p_f+\gamma_f$. Therefore, the
model consider  the threshold on  an end-to-end basis.  The motivation
for this second assumption will be given later in the section.

\vspace{-0.2cm}

\subsubsection{The parallel machine scheduling problem with setup time}

As for the model of Section~\ref{sec:CP}, the parallel machine
scheduling problem with setup time is model using interval variables
$jobs_j$, $\forall j \in \N$, and optional interval variables
$altJ_{j,m}$. The constraints used are the same and, therefore, are
not described in this section.

\paragraph{Cumulative constraints}
The model is also reinforced by considering the set of machines as a
cumulative resource of capacity $M$. Indeed, each job consumes one
unit of resource (one machine) during its execution and the total
capacity of the resource (total number of machines available) is
$M$. This is expressed using the global constraint
$cumulative$~\cite{globalConstraint}. 
\footnotesize
\begin{align}
 cumulative(\{(jobs_j,1)\, |\, \forall j \in \N\},M) 
\end{align}
\normalsize

\paragraph{Makespan modelling}

As for the previous model, the makespan of the scheduling is needed to
model machine disqualifications. The constraints presented in this
section concern the link between the makespan and the PMSP. A
constraint linking the makespan with the number of disqualifications
will be presented later in the paper.

Unlike the previous model, the makespan is modeled here as an interval
variable starting at time $0$ and spanning the execution of all
jobs. This is modelled using $span$ constraints. Introduced
in~\cite{Laborie09}, this constraint states that an executed
job must span over a set of other executed jobs by
synchronising its start date with the earliest start date of other
executed jobs and its end date with the latest end date. It is
expressed by the following constraints:
\footnotesize
\begin{align}
  span(C_{max}, \{ jobs_j \, | \, \forall j \in \N\}) \\
  st(C_{max}) = 0
\end{align}
\normalsize

In addition, the size of the interval has to be between the upper
and the lower bound on the makespan defined in
Remark~\ref{rem:boundCmax}.

\vspace{-0.2cm}

\subsubsection{Machine qualifications problem}

In this section, the model for the machine qualifications problem is
described. The main idea of the model is that, each time a job of
family $f$ is scheduled on a machine $m$, a {\it qualification
interval} of size $\gamma_{f}$ will occur right after. This interval
``models'' the fact that machine $m$ remains qualified for family $f$
until, at least, the end of the interval. To model this feature,
optional interval variable are used. Indeed, for each job $j$ and each
machine $m \in \M_{f(j)}$, an optional interval variable, $qual_{j,m}$, of size
$\gamma_{f(j)}$ and taking its value in
$\{0,\dots,T+\max_f{\gamma_f}\}$.  is created. Then, a variable
$qual_{j,m}$ will be present in the solution only if $altJ_{j,m}$ is
present and will start at the end of $altJ_{j,m}$. This is expressed
by the following set of constraints.
\footnotesize
\begin{align}
  & x(altJ_{j,m}) = x(qual_{j,m}) & \forall j \in \N,\ \forall m \in
                                     \M_{f(j)}\\
  & endAtStart(altJ_{j,m} , qual_{j,m}) & \forall j \in \N,\ \forall m \in
                                     \M_{f(j)}
\end{align}
\normalsize

Hence, a job of $f$ can only be scheduled on $m$ during a qualification
interval of $f$ on $m$. This is modeled using cumulative
functions. A cumulative function $Q_{f,m}$ counts, at each time $t$,
the number of qualification intervals for $(f,m)$ in which $t$ is.
If the number of qualification intervals for $(f,m)$ is greater than
$1$, then a job of $f$ can be scheduled on $m$. Otherwise, the number
of interval is zero and $m$ is disqualified for $f$. $Q_{f,m}$ is
expressed as: 

\[Q_{f,m} = pulse(0,\gamma_f+p_f,1) + \sum_{\substack{j\in \N \\
      f(j)=f}}\sum_{m\in \M_{f(j)}}   pulse(qual_{j,m},1) \]
Indeed, at the beginning of the scheduled, the machine is qualified
from time $0$ to $\gamma_f+p_f$. In addition, each time an interval variable
$qual_{j,m}$ is scheduled, $Q_{f(j),m}$ increases by one. Then, when a
job of $f$ is scheduled on $m$, $Q_{f,m}$ has to be 
greater than one and one can show that $Q_{f,m}$ is always smaller than
$n_f + 1$.
\footnotesize
\begin{align}
& alwaysIn(Q_{f(j),m}, altJ_{j,m}, 1, n_{f(j)}+1) & \forall j \in \N,\
\forall m \in \M_{f(j)}
\end{align}
\normalsize

\begin{example}[Example of cumulative function]
  Considering the instance of Example~\ref{ex:PTC}. The cumulative
  function $Q_{f_3,m_1}$ corresponding to
  Figure~\ref{subfig:exOptDisq} is described by
  Figure~\ref{fig:exCum}.
  
  \vspace{-0.5cm}
  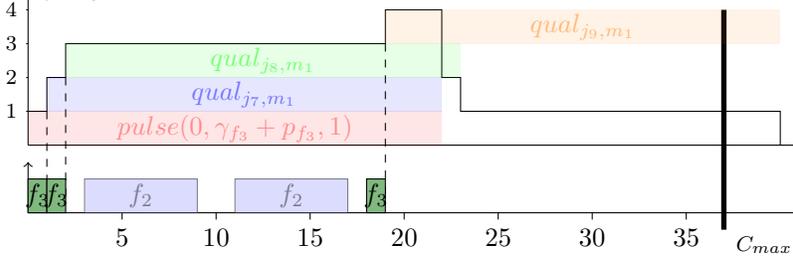
\begin{figure}[!htb]
    \begin{center}
      \begin{tikzpicture}[xscale=0.25,yscale=0.45]
        \node (O) at (0,0) {};
        \draw[->] (O.center) -- (41,0);
        \draw[->] (O.center) -- (0,4.5);

        \foreach \i in {1,2,3,4} {
          \draw (0,\i) -- (-0.1,\i) node[left] {\footnotesize \i};
        }
        
        \draw (0,1) -- (1,1) -- (1,2) -- (2,2) -- (2,3) -- (19,3) --
        (19,4) -- (22,4) -- (22,2) -- (23,2) -- (23,1) -- (40,1) --
        (40,0);

        \draw[line width=2pt]  (37,4) -- (37,  -2.5) node[below=0.2cm,
        right = 0.01cm]
        {\footnotesize $C_{max}$};
        
        \fill[red,opacity=0.1]       (0,0)      rectangle       (22,1)
        node[midway,opacity=0.4] {$pulse(0,\gamma_{f_3} + p_{f_3},1)$};
        \fill[blue,opacity=0.1]      (1,1)       rectangle      (22,2)
        node[midway,opacity=0.5] {$qual_{j_7,m_1}$}; 
        \fill[green,opacity=0.1]      (2,2)      rectangle      (23,3)
        node[midway,opacity=0.5] {$qual_{j_8,m_1}$}; 
        \fill[orange,opacity=0.1]     (19,3)      rectangle     (40,4)
        node[midway,opacity=0.5] {$qual_{j_9,m_1}$}; 

        \draw[dashed] (1,1) -- (1,-1.5) ;
        \draw[dashed] (2,2) -- (2,-1.5) ;
        \draw[dashed] (19,3) -- (19,-1.5) ;

        \begin{scope}[yshift=-2cm,yscale=2]
          \node (O) at (0,0) {};
          \draw[->] (O.center) -- (41,0);
          \draw[->] (O.center) -- (0, 0.75);
          \draw (5,0) -- (5,-0.1) node[below] {$5$} ;
          \draw (10,0) -- (10,-0.1) node[below] {$10$} ;
          \draw (15,0) -- (15,-0.1) node[below] {$15$} ;
          \draw (20,0) -- (20,-0.1) node[below] {$20$} ;
          \draw (25,0) -- (25,-0.1) node[below] {$25$} ;
          \draw (30,0) -- (30,-0.1) node[below] {$30$} ;
          \draw (35,0) -- (35,-0.1) node[below] {$35$} ;
    %      \draw[fill=red!15!]  (0,0.5) rectangle (9,1)
      %    node[midway,draw,fill=white,fill=white] {$f_1$}; 
        %  \draw[fill=red!15!]  (9,0.5) rectangle (18,1)
         % node[midway,draw,fill=white] {$f_1$}; 
      %    \draw[fill=red!15!]  (28,0.5) rectangle (37,1)
        %  node[midway,draw,fill=white] {$f_1$}; 
          \draw[fill=blue!30!,opacity = 0.45] (3,0) rectangle (9,0.5)
          node[midway] {$f_2$}; 
          \draw[fill=blue!30!, opacity = 0.45] (11,0) rectangle (17,0.5)
          node[midway] {$f_2$}; 
     %     \draw[pattern = dots]  (21,0.5) rectangle (27,1)
       %   node[midway,draw,fill=white] {$f_2$}; 
          \draw[fill=green!45!black!50!]  (0,0) rectangle (1,0.5) node[midway] {$f_3$};
          \draw[fill=green!45!black!50!] (1,0) rectangle (2,0.5) node[midway] {$f_3$};
          \draw[fill=green!45!black!50!] (18,0) rectangle (19,0.5) node[midway] {$f_3$};
    %      \draw (19,0.5) rectangle (20,1) node[midway] {$f_3$};
        \end{scope}
      \end{tikzpicture}
    \end{center}
    \vspace{-0.6cm}
    \caption{Example of cumulative function to model qualifications.}
    \label{fig:exCum}
   
    \vspace{-0.4cm}
  \end{figure}
  Each time a job of $f_3$ ends, the value of the function
  $Q_{f_3,m_1}$ increases by one and decreases when the qualification
  interval ends. While the value of $Q_{f_3,m_1}$ is greater than one,
  it is possible to schedule jobs of $f_3$ on $m_1$. Here,
  $Q_{f_3,m_1}$ is always greater than one for $t \in [0,C_{max})$
  meaning that $m_1$ remains qualified for $f_3$ at the end of the
  schedule. 
\end{example}

\paragraph{Machine disqualifications}

Another dummy optional interval variable set, $endQ_{f,m}$, is
introduced to check if a machine has been disqualified for a family
during the schedule. The variable is present in the final solution
only if the machine is still qualified at the end of the schedule. In
this case, the variable start at time $0$, ends at time $C_{max}+ p_f$
and the function $Q_{f,t}$ has to be greater than one during the whole
execution of job $endQ_{f,m}$ (otherwise, the machine has been
disqualified).
\footnotesize
\begin{align}
  &st(endQ_{f,m}) = 0 & \forall f \in \F,\ \forall m \in
                                     \M_{f}\\
  & endAtEnd(C_{max},endQ_{f,m},p_f)& \forall f \in \F,\ \forall m \in
                                         \M_{f}\\
  & alwaysIn(Q_{f,m} , endQ_{f,m}, 1, n_f+1) &  \forall f \in \F,\ \forall m \in
                                         \M_{f}
\end{align}
\normalsize
Note  that  this modelling  is  possible  because the  threshold  were
considered on an  end-to-end basis. Otherwise, the  use of constraints
{\it alwaysIn} is not possible.

\vspace{-0.2cm}
\subsubsection{Ordering constraints}

The following  sets of constraint  (partially) order variables  in the
solution. These (partial) ordering is  used to break symmetries in the
model. Recalling that it is assumed  that jobs of the same family have
consecutive  index in  $\N$. Then  constraints~\eqref{eq:orderStJob_1}
state   that   $jobs_{j-1}$  has   to   start   before  $jobs_j$   and
constraints~\eqref{eq:orderStJob_2} that the  maximum time lag between
these  jobs  is $\gamma_{f(j)}$.   Constraints~\eqref{eq:orderEtStJob}
order jobs  that cannot be executed  in parallel. Indeed, job  $j$ can
overlap  at most  $M_{f(j)} -1$.   Hence,  job $j-  M_{f(j)} $  cannot
overlap      job     $j$      and     has      to     end      before.
Constraints~\eqref{eq:orderStQual}   ensure  that   the  qualification
interval corresponding to job $j$  is separated from the qualification
interval  of $j-1$  by  at least  the duration  of  the job.  Finally,
constraints~\eqref{eq:orderStEtJob} model the fact  that, on a machine
$m$, jobs of  a same family are ordered, i.e.  smaller index scheduled
first.
\footnotesize
\begin{align}
  & startBeforeStart(jobs_{j-1},jobs_{j}) &
\forall j \in \N \ s.t. \ f(j) = f(j-1) \label{eq:orderStJob_1}\\
  & startBeforeSart(jobs_j,jobs_{j-1}, -\gamma_{f(j)})   &  \forall j \in \N \ s.t. \ f(j)
                                      = f(j-1) \label{eq:orderStJob_2}\\
  & endBeforeStart(jobs_{j - M_{f(j)}},jobs_j) &  \forall j \in \N \
                                                s.t. \ f(j) =
                                                f(j-M_{f(j)})\label{eq:orderEtStJob}\\ 
 & startBeforeStart(qual_{j - 1,m},qual_{j,m},p_{f(j)}) &  &
                                                           \nonumber\\
  & & \mathllap{\forall m \in
                                                        \M_{f(j)},\ \forall j
      \in \N \ s.t. \ f(j) = f(j-1)} \label{eq:orderStQual}\\
& endBeforeStart(altJ_{i,m},altJ_{j,m}) &  &
                                         \nonumber\\[-14pt]
  & &\mathllap{\forall m \in \M_{f(j)},\ \forall i < j \in \N \
                                     s.t. f(i)=f(j) }\label{eq:orderStEtJob}
\end{align}
\normalsize

\vspace{-0.4cm}

\subsubsection{Objective function}

The objective function is modeled using two integer variables:
$flowTime \in \{\underline{C_{max}},\dots,\overline{C_{max}}\}$ and
$qualified \in \{1,\dots, \sum_{f \in \F} M_f\}$. The expressions of these variables are
given below:
\footnotesize
\begin{align}
 & flowTime = \sum_{j \in \N} et(jobs_j) & \\
  & qualified = \sum_{f \in \F} \sum_{m \in \M_{f(j)}} x(endQ_{f,m})& 
\end{align}
\normalsize

Then, the objective is expressed as a sum, i.e. \FQ, or using the
lexicographical order, e.g. \QF.  Note that, in this model, the number
of machine qualified at the end of the schedule is maximized which is
equivalent to minimize the number of machine becoming disqualified
during the schedule.

\vspace{-0.2cm}
\subsubsection{Model size}

The number of variables of the model is at most $N\cdot (2M + 1) + M\cdot F + 3$ and
the number of constraints is at most $N^2\cdot M+ 4N + M \cdot (1 + 4N
+ 3F) + 6$.

\section{Experiments}
\label{sec:expe}

This section starts with the presentation of the instances used in
the experiments (Section~\ref{sec:instances}). Then, the general
framework of the experiments is described in~\ref{sec:framework}.
Finally, the three model presented in the paper are used to solve the
instances and the results are compared and analysed
(Section~\ref{sec:results}).

\subsection{Instance generation}
\label{sec:instances}

The benchmark instances used to perform our experiments are extracted
from~\cite{Nattaf2018C}. In this paper, 19 instance sets are generated
with different number of jobs ($N$), machines ($M$), family ($F$) and
qualification schemes.  Each of the instance sets is a group of 30
instances and are generated as follows.

In each generated instances, each family can be executed by at least
one machine and each machine is qualified to process at least one job
family. Furthermore, since short thresholds may lead to very quick
machine disqualifications, the time thresholds of job families are
chosen sufficiently large compared to their associated processing
times, i.e. $\max_{f \in \F}{p_f} \le \min_{f \in \F}{\gamma_f}$.
Then, to ensure diversity, each set of instances contains 10 instances
with small threshold (corresponding to duration needed to process one
to two jobs of another family than $f$), 10 with medium threshold
(two to three jobs) and 10 with large threshold (three to four jobs).
In addition, setup times are not chosen too large so that the risk of
disqualifying a machine due to a setup time insertion is
``acceptable'', i.e.  $\max_{f \in \F}{s_f} \le \min_{f \in \F}{p_f}$.

Table~\ref{tab:data} presents the parameters of the different instance
sets. In  the first rows, the  different number of jobs  $N$ is given,
number of  machines $M$ is  describedby the  second row and  number of
families $F$  is detailed  in the  third row.  Note that  each triplet
$(n,m,f)$  corresponds to  $30$ instances.  Among those  instances, at
least     $99,5\%$     are      feasible.      Indeed,     experiments
in~\cite{Nattaf2018C}  show  that only  one  60-job  instance and  two
70-job instances have  an unknown status. For all  other instances, at
least one of the algorithms presented in~\cite{Nattaf2018C} is able to
find a feasible solution.

 \begin{table}[!htb]
   \footnotesize
   \begin{center}
     \begin{tabularx}{\linewidth}{|P{0.5cm}|*{19}{C|}}
       \hline
       $N$ & \multicolumn{6}{c|}{20}&\multicolumn{6}{c|}{30}
       &40&50&\multicolumn{2}{c|}{60}&\multicolumn{3}{c|}{70}\nl 
       \hline
       $M$&\multicolumn{2}{c|}{3}&\multicolumn{4}{c|}{4}&\multicolumn{4}{c|}{3}&4&5&\multicolumn{5}{c|}{3}&\multicolumn{2}{c|}{4}\nl
       \hline
       $F$& 4&5&2&3&4&5&2&3&4&5&4&5&3&3&4&5&5&4&5\nl
       \hline
          \end{tabularx}
   \end{center}
   \caption{Instance characteristics}
   \label{tab:data}

    \vspace{-0.8cm}
 \end{table}

 The instances  generated are  relatively small compared  to industrial
instances.  However, due  to  the  complexity of  the  problem, it  is
important to first analyse and compare the results of the three models
described  in  this  paper.  Finding  good  solutions  for  industrial
instances is a  real challenge and is an  important research direction
for future work.

\subsection{Framework}
\label{sec:framework}

The experiment framework is defined so the following questions are
addressed:
\vspace{-0.25cm}
\begin{description}
\item[\normalfont \it Question 1.] Which  model is the best at finding
a feasible solution, proving the  optimality and/or finding good upper
bounds (especially when solving large instances)?
\item[\normalfont \it  Question 2.]  Does  the performance of  a model
change depending on the objective function or on the time limit?
\end{description}

The  models  are  implemented   using  IBM  Ilog  Optimization  Studio
12.8. That is CPLEX 12.8 for the  ILP model and CP Optimizer 12.8 for
CP models.   All the  experiments were  led on  a computer  running on
Ubuntu 16.04.5 with  32 GB of RAM and one  Intel Core i7-3930K 3.20GHz
processors (6  cores). Furthermore,  two time limits  are used  in the
experiments: $30$ and $600$ seconds.

Two heuristics  are used to find  solutions which are used  as a basis
for the  models. These heuristics  are called {\it  Scheduling Centric
Heuristic}        and         {\it        Qualification        Centric
Heuristic}~\cite{Nattaf2018C}. The  goal of the first  heuristic is to
minimize the  flow time  while the  second one  tries to  minimize the
number of disqualifications.

In  the  following  of  the  section   \IP  ,  \CP  and  \CPN  denotes
respectively the  \IP of section~\ref{sec:ILP}, the  previous CP model
described in  Section~\ref{sec:CP} and the  new CP model  detailled in
Section~\ref{sec:newCP}.  Furthermore, to  describe the performance of
the different models,  the following indicators are used  in the table
of  Section~\ref{sec:results}: $\%sol.$  described  the percentage  of
instances for  which feasible  solution is  found; $\%opt.$  shows the
percentage of  instances for which  the optimality is  proven; $\%vbs$
provides  the percentage  of  instances  for which  the  model is  the
virtual  best solver,  i.e. has  found the  best solution  compared to
others; $\#dis.$ gives the average number of disqualified machines and
finally, $obj.$ is used to show the average of the sum of the flowtime
and the number of disqualified machines.

In addition, a bold value in the table means that the corresponding
indicator has the best values among its row, i.e. compared to the one
of the others model.

\subsection{Comparison of the three models}
\label{sec:results}

This section aims at comparing the results of the three models. First,
the results are described for the {\it tight time limit}, i.e. $30$
seconds. Then, the results with the $600$-seconds time limits are
given.

\subsubsection{$30$-seconds time limit}

 \vspace{-0.3cm}

\paragraph{Minimizing the number of disqualifications over the
  flow time} 
Table~\ref{tab:QF30} gives indicators for the three models solved
using the \QF objective with $30$-seconds time limit.

\begin{table}[htbp]
  \centering
  \begin{tabularx}{\linewidth}{c|CCCC|CCCC|CCCC}
    \toprule
    \ttit{}& \multicolumn{4}{c}{\IP }  & \multicolumn{4}{c}{\CP} & \multicolumn{4}{c}{\CPN} \\
    \cmidrule(lr){2-5}     \cmidrule(lr){6-9} \cmidrule(lr){10-13}
    ~$N$~ & $\%sol .$& $\%opt.$ & $\%vbs$ & \ttit{$\#dis.$} & $\%sol .$& $\%opt.$
    & $\%vbs$ & \ttit{$\#dis.$} & $\%sol .$& $\%opt.$ & $\%vbs$ & \ttit{$\#dis.$} \\ 
    \midrule
    20 & \bf 100 & 54.4 & 55.6 & 1.1 & \bf  100 & 69.4 & 86.1 &\bf 0.6
    & \bf 100 & \bf 82.2  &\bf 90.6 & \bf 0.6 \nl
    30 & 97.2 & 21.7 & 23.3 & 3.1 &\bf  99.4  & 51.1 & 59.4 &    1.4 & 98.9      &\bf 56.7 &\bf 71.1 &\bf 1.2 \nl
    40 & \bf 100 & 23.3 & 26.7 & 0.9 & \bf 100 & 63.3 & 63.3 &    0.6
    & \bf 100 &\bf 83.3 &\bf 90 &\bf 0.2 \nl
    50 & \bf 100 &  0 & 6.7  & 2.9 & \bf  100 & 33.3 & 36.7 &    1.4
    & \bf 100 &\bf 56.7 &\bf 73.3 &\bf 0.8 \nl
    60 &  88.3 &  0 & 0  & 7.5 & \bf 90  & 8.3  & 33.3 &    3.4 & \bf 90      &\bf 21.7 &\bf 56.7 &\bf 2.8 \nl
    70 &  86.7 &  0 & 0  & 9.5 & \bf  91.1  & 4.4  & 41.1 &    5 & \bf 91.1      &\bf 15.6 &\bf 52.2 &\bf 4.1 \nl     
    \bottomrule
  \end{tabularx}
  \caption{Lexicographic minimization of the disqualified machines and the flowtime within 30 seconds.}
  \label{tab:QF30}  
 \vspace{-0.8cm}
\end{table}

Table~\ref{tab:QF30} shows that the \IP finds less feasible
solutions than the CP models. Furthermore, the \IP does not scale well
for large instances. Indeed, the \IP is never the VBS and its average
number of disqualified machines is very high for the largest instances
compared to the CP models.

On the other hand, the \CPN obtains better results than the
\CP. Indeed, the percentage of proof of optimality is higher with
\CPN.  The model is also more often the VBS regardless of the instance
size.  Furthermore, the difference between the average numbers of
qualified machines of both model increases with the instance
size. This shows that the \CPN scales better than the \CP.

\paragraph{Minimizing the flow time over the number of disqualifications}

Table~\ref{tab:FQ30} gives indicators for the three models solved
using the \FQ objective with $30$-seconds time limit.

\begin{table}[htbp]
  \centering
  \begin{tabularx}{\linewidth}{c|CCCC|CCCC|CCCC}
    \toprule
    \ttit{}& \multicolumn{4}{c}{\IP }  & \multicolumn{4}{c}{\CP} & \multicolumn{4}{c}{\CPN} \\
    \cmidrule(lr){2-5}     \cmidrule(lr){6-9} \cmidrule(lr){10-13}
    ~$N$~ & $\%sol.$ & $\%opt.$ & $\%vbs$ & \ttit{$obj.$} & $\%sol .$& $\%opt.$
    & $\%vbs$ & \ttit{$obj.$} & $\%sol .$& $\%opt.$ & $\%vbs$ & \ttit{$obj.$} \\ 
    \midrule     
    20 & \bf 100 & \bf 96.7 & \bf 97.8 & \bf 334.7 & \bf 100 & 0 &
    87.8     & 334.8      & \bf 100 & 65.6 & 90  & \bf 334.7  \nl 
    30 & 97.8      & \bf 69.4 & \bf 72.2 & 782.5     & \bf 99.4  & 0 &
    71.1     & 770      & 98.9      & 24.4 & 57.8  & \bf  766.8 \nl 
    40 & \bf 100 & \bf 90 & 90     & 1536    & \bf 100 & 0 &  93.3
    & 1530     & \bf 100 & 60 & \bf100 & \bf 1529 \nl 
    50 & \bf 100 & \bf 60 & 70     & 2265    & \bf 100 & 0 & \bf
    76.7 & 2159     & \bf 100 & 10 & 73.3  & \bf 2151 \nl 
    60 & 88.3      & \bf 5  & 8.3      & 3228    & \bf 90  & 0 & \bf
    50 & \bf 2792 & \bf 90  & 0  & 36.7  & 2805     \nl 
    70 & 86.7      & \bf 4.4  & 5.6      & 4256    & 90      & 0 &
    \bf 52.2 & 3583     & \bf 91.1  & 0  & 43.3  & \bf 3562 \nl 
    \bottomrule
  \end{tabularx}
  
  \caption{Weighted sum minimization of the flowtime and number of disqualified machines within 30 seconds.}
  \label{tab:FQ30}
  
 \vspace{-0.9cm}
\end{table}

Table~\ref{tab:FQ30} shows that the \IP is more
competitive than when the priority is given to the number of
disqualifications. Indeed, despite the fact that it finds a little less
feasible solutions than the CP Models, it is better at proving the
optimality of its solution. However, the \IP does not scale well as
shown by the high objective values for the largest instances. 

On the other hand, the \CP is the most efficient for finding good
upper bounds, but completely fails at proving the optimality of its
solution.  The \CPN proves optimality less often than the
\IP. However, it is only slightly dominated by the \CP in terms of 
being the VBS.  However, the \CPN still have the lowest objective
values.

\subsubsection{$600$-seconds time limit}

Table~\ref{tab:loose} gives indicators for the three models solved
using both lexicographic and weighted sum minimization with
$30$-seconds and $600$-seconds time limit. Only challenging instances
with 60 jobs are used to save computation time. 

For the lexicographic minimization, the \CPN confirms its predominance.
For all three models, the percentages of solved instances remain
constant, the percentages of optimality proof only slightly increase,
and the average numbers of disqualified machines significantly
decrease. 

For the weighted sum minimization, the \IP becomes the best model. The
percentages of solved instances and optimality proof significantly
improve and the model often becomes the VBS. Nevertheless, the \CPN
model has the best average objective. 

Most of the times, the low improvements of the number of solved
instances or optimality proofs suggest that the solvers is subject to
thrashing and therefore cannot diversify the search. 

 \vspace{-0.4cm}
\begin{table}[htbp]
  \centering
  \begin{tabularx}{\linewidth}{c|CCCC|CCCC|CCCC}
    \toprule
    \ttit{}& \multicolumn{4}{c}{\IP }  & \multicolumn{4}{c}{\CP} & \multicolumn{4}{c}{\CPN} \\
    \cmidrule(lr){2-5}     \cmidrule(lr){6-9} \cmidrule(lr){10-13}

    ~$t$~ & $\%sol.$ & $\%opt.$ & $\%vbs$ & \ttit{$\#dis.$} & $\%sol
    .$ & $\%opt.$ & $\%vbs$   & \ttit{$\#dis.$} & $\%sol .$ & $\%opt.$
    & $\%vbs$  & \ttit{$\#dis.$} \\  
    \midrule     
    30s   & 88.3     & 0        & 0       & 7.5             & \bf 90    & 8.3      & 33.3      & 3.4             & \bf 90    & \bf 21.7 & \bf 56.7 & \bf 2.8 \nl
    600s   & \bf 90   & 0        & 3.3     & 4.2             & \bf 90    & 11.7     & 28.3      & 2.9             & \bf 90    & \bf 23.3 & \bf 61.7 & \bf  2.2 \nl
    \midrule
    ~$t$~ & $\%sol.$ & $\%opt.$ & $\%vbs$ & \ttit{$obj.$}   & $\%sol
    .$ & $\%opt.$ & $\%vbs$   & \ttit{$obj.$}   & $\%sol .$ & $\%opt.$
    & $\%vbs$  & \ttit{$obj.$}   \\  
    \midrule     
    30s   & 88.3     & \bf 5    & 8.3     & 3228            & \bf 90    & 0        & \bf    50 & \bf 2792        & \bf 90    & 0        & 36.7     & 2805     \nl 
    600s  & \bf 98.3 & \bf 55   & \bf 75  & 2873          & 90        & 0        & 33.3      & 2755          & 90        & 0        & 33.3     & \bf 2744 \nl

    \bottomrule
  \end{tabularx}
  \caption{Weighted sum and leximin minimization over instances of 60 jobs within 600 seconds.}
  \label{tab:loose}

   \vspace{-1.5cm}
\end{table}

\vspace{-0.1cm}
\section{Conclusions and further work}
\vspace{-0.1cm}

A parallel machine scheduling problem was studied where some
Advanced Process Control constraints are integrated: minimal time
constraints between jobs of the same family to be processed on a
qualified machine to avoid losing the qualification. Two criteria to
minimize are considered: the sum of completion times and the number of
disqualifications.

For this problem, a new CP model was proposed. This model improves the
modelling of machine disqualifications. Indeed, when the number of
disqualifications is prioritized, this model is better than the
existing methods (\IP and \CP) in terms of objective value and in
terms of optimality proof. However, when the flow time is prioritized,
the performance of the model is less impresive. In this case, the \CP
tends to have better performance for small-time limit and the \IP
performs better in case of larger time limit.

Experiment results show that a good CP model needs to make some
improvements on the modelling and/or the solving of the parallel
machine scheduling problem with the flow time objective. Interesting
research directions include the improvement of variable bounds,
specially the makespan. It also includes the study of good relaxations 
of the problem to enhance the performance of constraint programming
models.

Another relevant research perspective consists in scheduling jobs on a
longer time horizon, where  lost qualifications could be automatically
recovered  after  a  given qualification  procedure.  Qualification
procedures, requiring time on machines, would then also be scheduled.

\bibliographystyle{splncs04}
\bibliography{biblioCPAIOR}

\begin{thebibliography}{10}
\providecommand{\url}[1]{\texttt{#1}}
\providecommand{\urlprefix}{URL }
\providecommand{\doi}[1]{https://doi.org/#1}

\bibitem{globalConstraint}
Beldiceanu, N., Carlsson, M., Rampon, J.X.: Global constraint catalog (revision
  a)  (01 2012)

\bibitem{Cai2011}
Cai, Y., Kutanoglu, E., Hasenbein, J., Qin, J.: Single-machine scheduling with
  advanced process control constraints. Journal of Scheduling  \textbf{15}(2),
  165--179 (Apr 2012). \doi{10.1007/s10951-010-0215-8},
  \url{https://doi.org/10.1007/s10951-010-0215-8}

\bibitem{Jedidi2011}
Jedidi, N., Sallagoity, P., Roussy, A., Dauzere-Peres, S.: Feedforward
  run-to-run control for reduced parametric transistor variation in cmos logic
  0.13 {$\mu{\rm m}$} technology. IEEE Transactions on Semiconductor
  Manufacturing  \textbf{24}(2),  273 --279 (2011)

\bibitem{Laborie08}
Laborie, P., Rogerie, J.: Reasoning with conditional time-intervals. In:
  Proceedings of the Twenty-First International Florida Artificial Intelligence
  Research Society Conference, May 15-17, 2008, Coconut Grove, Florida, {USA}.
  pp. 555--560 (2008),
  \url{http://www.aaai.org/Library/FLAIRS/2008/flairs08-126.php}

\bibitem{Laborie09}
Laborie, P., Rogerie, J., Shaw, P., Vil{\'{\i}}m, P.: Reasoning with
  conditional time-intervals. part {II:} an algebraical model for resources.
  In: Proceedings of the Twenty-Second International Florida Artificial
  Intelligence Research Society Conference, May 19-21, 2009, Sanibel Island,
  Florida, {USA} (2009),
  \url{http://aaai.org/ocs/index.php/FLAIRS/2009/paper/view/60}

\bibitem{Li2008}
Li, L., Qiao, F.: The impact of the qual-run requirements of {APC} on the
  scheduling performance in semiconductor manufacturing. In: Proceedings of
  2008 {IEEE} International Conference on Automation Science and
  Engineering(CASE). pp. 242--246 (2008)

\bibitem{Moench2011}
Moench, L., Fowler, J.W., Dauz\`ere-P\'er\`es, S., Mason, S.J., Rose, O.: A
  survey of problems, solution techniques, and future challenges in scheduling
  semiconductor manufacturing operations. Journal of Scheduling pp. 1--17
  (2011), \url{http://dx.doi.org/10.1007/s10951-010-0222-9},
  10.1007/s10951-010-0222-9

\bibitem{Moyne2000}
Moyne, J., del Castillo, E., Hurwitz, A.M.: {Run-to-Run} Control in
  Semiconductor Manufacturing. {CRC} Press, 1 edn. (2000)

\bibitem{Musacchio1997}
Musacchio, J., Rangan, S., Spanos, C., Poolla, K.: On the utility of run to run
  control in semiconductor manufacturing. In: Proceedings of 1997 {IEEE}
  International Symposium on Semiconductor Manufacturing Conference. pp. 9--12
  (1997)

\bibitem{Nattaf2018C}
Nattaf, M., Dauzère-Pérès, S., Yugma, C., Wu, C.H.: Parallel machine
  scheduling with time constraints on machine qualifications, {M}anuscript
  submitted for publication.

\bibitem{Nattaf2018R}
Nattaf, M., Obeid, A., Dauzère-Pérès, S., Yugma, C.: Méthodes de
  résolution pour l’ordonnancement de familles de tâches sur machines
  parallèles et avec contraintes de temps. In: 19ème édition du congrès
  annuel de la Société Française de Recherche Opérationnelle et d'Aide à
  la Décision, ROADEF2018

\bibitem{Obeid2014}
Obeid, A., Dauz{\`e}re-P{\'e}r{\`e}s, S., Yugma, C.: Scheduling job families on
  non-identical parallel machines with time constraints. Annals of Operations
  Research  \textbf{213}(1),  221--234 (Feb 2014).
  \doi{10.1007/s10479-012-1107-4}

\bibitem{Tan2015}
Tan, F., Pan, T., Li, Z., Chen, S.: Survey on run-to-run control algorithms in
  high-mix semiconductor manufacturing processes. IEEE Transactions on
  Industrial Informatics  \textbf{11}(6),  1435--1444 (2015)

\bibitem{Wolf2009}
Wolf, A.: Constraint-based task scheduling with sequence dependent setup times,
  time windows and breaks. In: GI Jahrestagung (2009)

\bibitem{Yugma2015}
Yugma, C., Blue, J., Dauz{\`e}re-P{\'e}r{\`e}s, S., Obeid, A.: Integration of
  scheduling and advanced process control in semiconductor manufacturing:
  review and outlook. Journal of Scheduling  \textbf{18}(2),  195--205 (Apr
  2015). \doi{10.1007/s10951-014-0381-1},
  \url{https://doi.org/10.1007/s10951-014-0381-1}

\end{thebibliography}
\end{document}